# Is Betelgeuse Really Rotating? Synthetic ALMA Observations of Large-scale Convection in 3D Simulations of Red Supergiants

Jing-Ze Ma (马竟泽)[1], Andrea Chiavassa[1,2], Selma E. de Mink[1,3], Ruggero Valli[1], Stephen Justham[1,4,5], and Bernd Freytag[6]
[1] Max Planck Institute for Astrophysics, Karl-Schwarzschild-Str. 1, 85748 Garching, Germany; jingze@mpa-garching.mpg.de
[2] Université Côte d'Azur, Observatoire de la Côte d'Azur, CNRS, Lagrange, CS 34229, Nice, France
[3] Anton Pannekoek Institute for Astronomy and GRAPPA, University of Amsterdam, NL-1090 GE Amsterdam, The Netherlands
[4] School of Astronomy & Space Science, University of the Chinese Academy of Sciences, Beijing 100012, People's Republic of China
[5] National Astronomical Observatories, Chinese Academy of Sciences, Beijing 100012, People's Republic of China
[6] Theoretical Astrophysics, Department of Physics and Astronomy, Uppsala University, Box 516, SE-751 20 Uppsala, Sweden
Received 2023 November 23; revised 2024 January 25; accepted 2024 January 25; published 2024 February 16

## Abstract

The evolved stages of massive stars are poorly understood, but invaluable constraints can be derived from spatially resolved observations of nearby red supergiants, such as Betelgeuse. Atacama Large Millimeter/submillimeter Array (ALMA) observations of Betelgeuse showing a dipolar velocity field have been interpreted as evidence for a projected rotation rate of about 5 km s$^{-1}$. This is 2 orders of magnitude larger than predicted by single-star evolution, which led to suggestions that Betelgeuse is a binary merger. We propose instead that large-scale convective motions can mimic rotation, especially if they are only partially resolved. We support this claim with 3D CO5BOLD simulations of nonrotating red supergiants that we postprocessed to predict ALMA images and SiO spectra. We show that our synthetic radial velocity maps have a 90% chance of being falsely interpreted as evidence for a projected rotation rate of 2 km s$^{-1}$ or larger for our fiducial simulation. We conclude that we need at least another ALMA observation to firmly establish whether Betelgeuse is indeed rapidly rotating. Such observations would also provide insight into the role of angular momentum and binary interaction in the late evolutionary stages. The data will further probe the structure and complex physical processes in the atmospheres of red supergiants, which are immediate progenitors of supernovae and are believed to be essential in the formation of gravitational-wave sources.

*Unified Astronomy Thesaurus concepts:* Red supergiant stars (1375); Stellar convection envelopes (299); Stellar rotation (1629); Astronomical simulations (1857); Hydrodynamics (1963); Radiative transfer (1335); Supergiant stars (1661); Submillimeter astronomy (1647)

*Supporting material:* animation

## 1. Introduction

Cool evolved stars are not expected to be rotating fast, at least not at their surfaces. As the stars evolve, their envelopes expand by 1–2 orders of magnitude. The outer layers thus slow down as a result of angular momentum conservation and may be further reduced by, e.g., mass loss due to stellar winds (e.g., Maeder & Meynet 2000; Smith 2014), possibly inward convective transport of angular momentum (e.g., Brun & Toomre 2002; Brun & Palacios 2009), and magnetic braking (Mestel 1968). The theory of single-star evolution therefore predicts slow surface rotation rates, less than about 1 km s$^{-1}$ for stars at the tip of the red giant branch (e.g., Privitera et al. 2016a) and less than about 0.1 km s$^{-1}$ for red supergiants (RSGs; Wheeler et al. 2017; Chatzopoulos et al. 2020), which are the cool giant descendants of massive stars.

Despite theoretical expectations, cool stars with rotation rates exceeding these predictions have been observed across the Hertzsprung–Russell diagram. These include several hundred red giants, about 1% of the total population of red giants (e.g., Patton et al. 2024, and references therein), and a few asymptotic giant branch (AGB) stars (Barnbaum et al. 1995; Vlemmings et al. 2018; Brunner et al. 2019; Nhung et al. 2021, 2023). For RSGs, so far only one has been claimed to rotate rapidly: α Orionis, better known as Betelgeuse (Uitenbroek et al. 1998; Harper & Brown 2006; Kervella et al. 2018), which recently has drawn wide attention after the sudden Great Dimming (Guinan et al. 2019; Montargès et al. 2021) and subsequent rebrightening (Guinan et al. 2020; Dupree et al. 2022).

Betelgeuse, being one of the closest RSGs to Earth, is one of the few stars that can be spatially resolved and has therefore been a target of interferometric studies for over a century (Michelson & Pease 1921). Recently, the Atacama Large Millimeter/submillimeter Array (ALMA) provided unprecedented maps of the molecular envelope (Kervella et al. 2018, hereafter K18; right-hand panels of Figure 2). The surface radial velocity map shows a remarkably clear dipolar structure; half of the visible hemisphere of the star shows a blueshift, and the other half shows a redshift of several km s$^{-1}$.

A natural explanation of such a dipolar velocity field is stellar rotation, as noted by K18. They inferred a projected equatorial velocity of $v \sin i = 5.47 \pm 0.25$ km s$^{-1}$. They compared the results with earlier measurements using the Hubble Space Telescope (HST) probing the chromosphere (Uitenbroek et al. 1998; Harper & Brown 2006) and argued that both the ALMA and HST data are consistent with the interpretation that Betelgeuse is fast-rotating. The fast rotation rate inferred for Betelgeuse is surprising in light of the predictions of single-star models, as illustrated in Figure 1 and Appendix A for details.

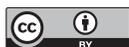

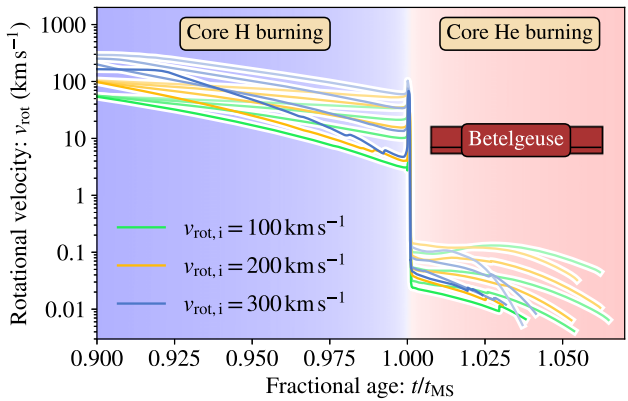

**Figure 1.** The equatorial rotational velocity inferred for Betelgeuse (K18) is 2–3 orders of magnitude higher than single-star models predict. To illustrate this point, we show MESA predictions for the rotational velocities of massive stars (12, 15, 18, and 21 $M_\odot$, shown in lines with increasing color saturation) with typical initial rotation rates as a function of their fractional age, i.e., time divided by their main-sequence lifetime. The steep drop in rotation rates marks the rapid expansion as stars transition from the core H-burning phase (blue background) to the core He-burning RSGs (red background). The red box labeled "Betelgeuse" shows the measured rotational velocity of the star, where the bottom, middle, and top black lines correspond to different assumptions for the inclination with respect to the observer. See Appendix A for details.

Binary star evolution has been proposed as an explanation for Betelgeuse's high rotation rate, in particular the merger with a lower-mass companion (Wheeler et al. 2017; Chatzopoulos et al. 2020; Sullivan et al. 2020; Shiber et al. 2023). This may seem like an exotic explanation, but massive stars often interact with close companions (Sana et al. 2012). As a consequence, stellar mergers are expected to be common (de Mink et al. 2014; Kochanek et al. 2014). Zapartas et al. (2019) estimated that as many as one-third of RSGs experience a stellar merger before they reach core collapse. Rui & Fuller (2021) identified two dozen red giants that are possible merger products based on their asteroseismological signatures. For red giants, the engulfment of planets has also been proposed as an explanation for their rapid rotation (Carlberg et al. 2012; Privitera et al. 2016b; Gaulme et al. 2020; Lau et al. 2022).

Establishing whether Betelgeuse is indeed rotating is of vital importance to better understand its evolutionary history, the possible role of binary interaction, and the physics of the evolved stages of massive stars in general (see Wheeler & Chatzopoulos 2023, for a review). Unfortunately, accurate and reliable measurements of rotation rates for red (super)giants are challenging.

The first complicating factor concerns the high velocities expected for convective flows at the photosphere. These may be as high as 20 km s$^{-1}$, as shown in different 3D simulations (Kravchenko et al. 2019; Antoni & Quataert 2022; Goldberg et al. 2022) as well as spectroscopic (Lobel & Dupree 2000; Josselin & Plez 2007) and optical interferometric observations (Ohnaka et al. 2009, 2011, 2013, 2017). This is 2 orders of magnitude larger than the predicted rotational velocities and 4 times larger than the rotational velocity inferred for Betelgeuse by K18. How the turbulent velocity field affects the measurement of the rotation rate is not yet well understood.

A second complication concerns the large sizes expected for the convective cells at the surface, which may span a significant fraction of the radius (Schwarzschild 1975). Only a few of them will be present at the surface at any given time, as also suggested by spectropolarimetric (López Ariste et al. 2018, 2022) and optical interferometric observations (e.g., Haubois et al. 2009; Norris et al. 2021). If, by chance, one very large cell or a group of cells move toward the observer while others move away, this can result in a dipolar velocity field even for a *nonrotating* star.

The central question motivating our current study is, "Can a *nonrotating* red (super)giant be mistaken to be a rapid rotator?" For our study, we use existing 3D radiation hydrodynamic simulations of RSGs with properties similar to Betelgeuse. We develop a new postprocessing package to solve the radiative transfer equations and make direct predictions for ALMA observables that we compare with observations of Betelgeuse. We quantify how fast a nonrotating star can appear to be rotating and how likely it is to obtain spurious measurements of high rotation. We conclude that, to firmly establish whether Betelgeuse is rotating rapidly, additional epochs of ALMA observations are needed, preferably with higher spatial resolution.

## 2. Method

### 2.1. 3D Simulations of RSG Envelope with CO5BOLD

To assess whether a nonrotating RSG can show a dipolar radial velocity map in the ALMA band, we need global 3D RSG models that simulate the multiscale convection of the full star. So far this is only possible with the CO5BOLD models (Freytag et al. 2012), since other 3D RSG models do not simulate the whole $4\pi$ sphere (Goldberg et al. 2022). The CO5BOLD RSG simulations have been extensively used to interpret spectrophotometric, interferometric, and astrometric observations, especially in the context of Betelgeuse (e.g., Chiavassa et al. 2009, 2010; Montargès et al. 2014, 2016; Kravchenko et al. 2021). The code numerically integrates the nonlinear compressible hydrodynamic equations, coupled with a short-characteristics scheme for radiation transport (Freytag et al. 2012) such that it accounts for the heating and cooling effect of the radiation flux. The global simulations that we use in this work adopt the "star-in-a-box" setup by simulating the outer part of the convective envelope with mass $M_{\rm env}$ on an equidistant Cartesian grid. The interior is replaced by an artificial central region providing a luminosity source with a drag force to damp the velocity. A gravitational potential is imposed, set by the central mass $M_{\rm pot}$. For the gravity experienced by the simulation, the stellar mass is $M_{\rm pot}$ because the self-gravity of the envelope is neglected. However, for actual stars, the total stellar mass would be $M_{\rm pot} + M_{\rm env}$. For detailed descriptions of the general setup, see Chiavassa et al. (2011a), Freytag et al. (2012), and Chiavassa et al. (2024).

We use the 3D radiation hydrodynamic simulations of RSG envelopes presented in Ahmad et al. (2023, and references therein). We chose their model `st35gm04n37` for the discussion presented in the main text of this paper (hereafter referred to as our fiducial model or model A). The assumed surface gravity and effective temperature in this model are close to the values observed for Betelgeuse (see Table 2 in Appendix D), but the model is not a perfect match. Therefore, we also analyze an alternative model, `st35gm03n020` from Ahmad et al. (2023), hereafter model B, which is more massive. The mass adopted in this model is closer to the inferred mass of Betelgeuse, but it may nevertheless be less appropriate than model A. Arroyo-Torres et al. (2015) show that model B has less extended atmosphere than observed. This is probably because the radiation pressure is neglected in the simulations (Chiavassa 2022). The lower mass and lower surface gravity in model A partly compensate for this. Recent studies therefore used model A to compare with observations of other RSGs that are similar to Betelgeuse (Kravchenko et al. 2019; Climent et al. 2020; Chiavassa et al. 2022). We only present model A in the main text for conciseness but refer the reader to Appendix C.4 for a discussion of the limitations and Appendix D for an analysis of model B. The quantitative results differ, but for both models, we find the same main conclusion that convective motions can mimic rapid rotation at the km s$^{-1}$ level.

### 2.2. Synthetic ALMA Images

We postprocess the 3D simulations to create synthetic observations for ALMA. The steps toward the synthetic images are as follows.

1. Calculate the abundances of atoms and molecules in each cell of 3D simulations to get emissivity and opacity.
2. Solve the radiative transfer equations to obtain images of intensity maps.
3. Convolve the resulting images with the ALMA beam.

Here we describe the main assumptions. Details and tests for each step are provided in Appendix C. The postprocessing package, animations, and data behind the figures are publicly available at Zenodo: doi:10.5281/zenodo.10199936.

To directly compare with the SiO line spectra observed by ALMA, we need to calculate the intensity from SiO emission and absorption. For the chemical abundances of the relevant species (SiO molecules, the electrons, and atomic H), we use the equilibrium chemistry code FASTCHEM2[7] (Stock et al. 2018, 2022). FASTCHEM2 has been widely adopted and tested against observations of exoplanetary atmospheres, in particular for ultrahot Jupiters, where the dayside conditions are similar to cool stars (e.g., Kitzmann et al. 2018, 2023). Here, we present the first application of FASTCHEM2 to stellar atmospheres. Chemical equilibrium may not be a good approximation for the shock-dominated atmosphere considered here (e.g., Cherchneff 2006). However, since this study mainly focuses on the velocity field rather than the abundances, the impacts on the main results are expected to be limited.

To synthesize the intensity map, we numerically integrate the radiative transfer equation on a Cartesian grid. We take into account the continuum free–free opacity and Doppler-shifted $^{28}$SiO lines (vibrational level $\nu = 2$, rotational transition $J = 8-7$ as observed by ALMA). Detailed equations and opacity sources can be found in Appendix C.2. Since one actual ALMA image of Betelgeuse only needs less than 1 hr to take (K18), the surface motions are approximately frozen during the exposure time of one image. Therefore, no extra integration in time is needed to produce synthetic maps from simulation snapshots.

We assume that the simulated star is located at a distance away from the observers such that the radio photospheric radius approximately matches the ALMA observations in K18 (for the associated limitations, see Appendix C.4). We interpolate the intensities onto a $101^2$ pixel grid, such that both the pixel number and the angular coverage are identical to ALMA observations. We then convolve the intensity map with the ALMA beam (FWHM of 18 mas). We use the convolved continuum intensity map to calculate the radius of the radio photosphere $R_{\rm radio} = \sqrt{\int I(r,\theta) r dr d\theta / (\pi I_{\rm star})}$, where the integral is performed over the 2D image (Section 3.3.1 in K18). Here, $I$ is the continuum intensity as a function of the polar coordinate $(r, \theta)$ on the 2D image, measured in the spectral window centered at 343.38 GHz as done in the ALMA observations. Its value at the photocenter is denoted as $I_{\rm star}$.

### 2.3. Analysis of the Radial Velocity Map

To compare our simulations with the observed radial velocity maps, we apply similar analysis methods as in K18 to our generated synthetic observations. We use the continuum-subtracted intensities to identify the line in each pixel, fit a Gaussian profile to the line using least-squares fitting, and obtain the radial velocity shift from the central value of the Gaussian. The apparent systematic velocity $v_{\rm sys}$ is obtained from the radial velocity shift of the integrated line (over the region up to 30 mas from the center for the absorption line and 30–50 mas for the emission line as in Section 3.3.2 in K18).

We measure the apparent $v \sin i$ by fitting the projected radial velocity map of a rigidly rotating sphere to the $v_{\rm sys}$-subtracted radial velocity map within $R_{\rm radio}$ (Section 3.5 in K18; however, see a more detailed discussion in Appendix C.3 for the choice of the radius adopted). Since both the observing time and the orientation of the star with respect to the observer are arbitrary, we repeat this procedure for six faces of the Cartesian box and every snapshot of the simulations. Throughout the main text, we use the synthetic radial velocity maps derived from emission lines and not absorption lines, in accordance with the actual analysis of K18.

## 3. Results

### 3.1. Selected Mock Observations Compared to Actual ALMA Observations

In Figure 2, we show a selected example of a synthetic ALMA image from our nonrotating RSG simulation and compare it with the ALMA observations of Betelgeuse in K18. The left-hand panels ((a) and (d)) show the original simulation. The middle panels ((b) and (e)) show our synthetic observations after convolution with the ALMA beam. The right-hand panels ((c) and (f)) show the actual ALMA observations (K18). In the top row, we show the simulated and observed continuum intensity maps. In the bottom row, we show the radial velocity maps measured from the Doppler shifts of the lines.

The unconvolved intensity map at these wavelengths probes the atmospheric layers of the star, which are highly asymmetric (Figure 2(a)). The convective motions of $\pm 30$ km s$^{-1}$ can be seen in the radial velocity map of the original simulations (Figure 2(d)). Both the cell size and the peak convective velocity are consistent with analytical estimates (Appendix B) and other 3D simulations of RSGs (Kravchenko et al. 2019; Antoni & Quataert 2022; Goldberg et al. 2022).

The ALMA beam used in the settings by K18 is about 30% of the diameter of the radio photosphere. This means that any sharp features on the surface will be smoothed out. The effect of this can be seen when comparing the left and middle panels in Figure 2. For the intensity map, a single hot spot emerges (Figure 2(b)), similar to what is observed by ALMA (Figure 2(c)). Such a

---
[7] https://github.com/exoclime/FastChem

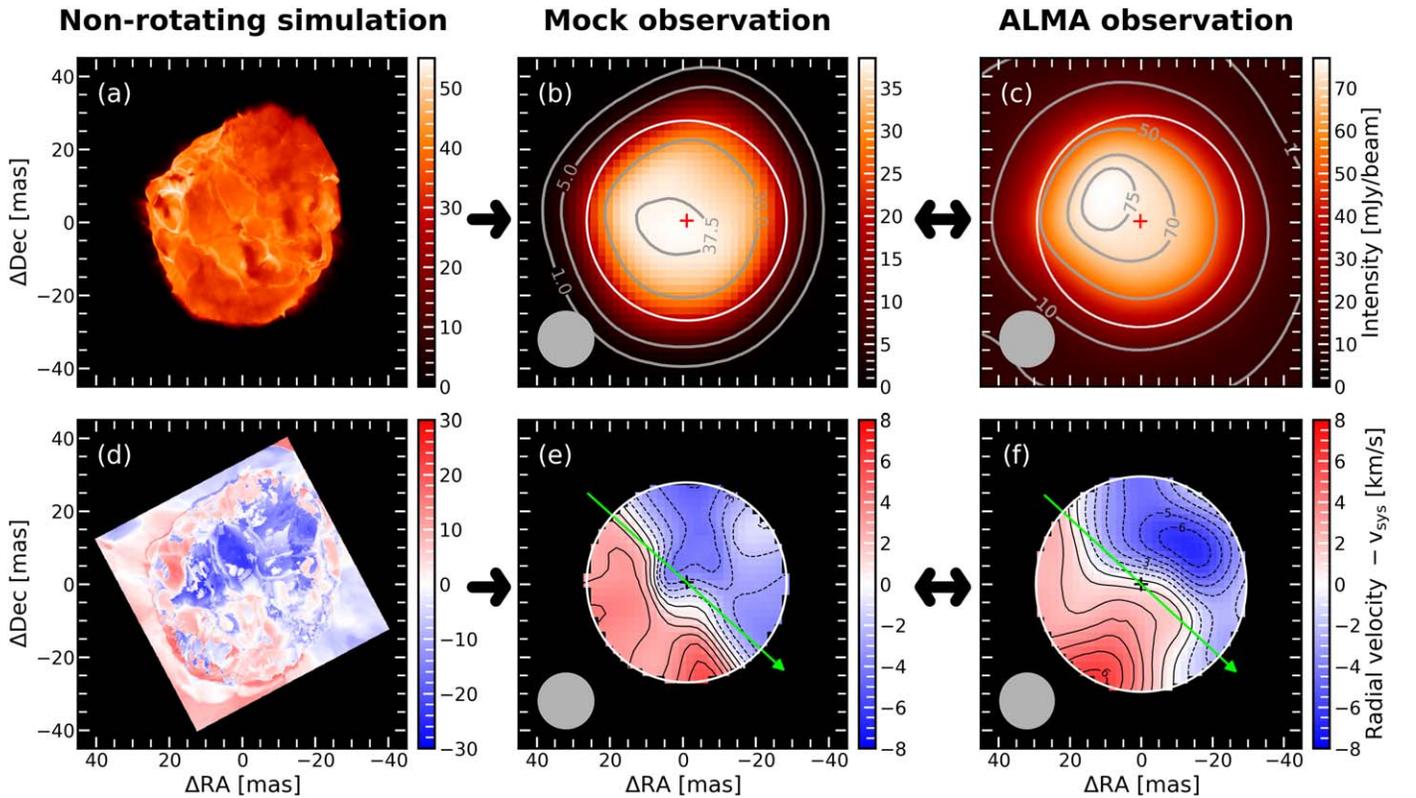

**Figure 2.** A direct comparison of a nonrotating RSG simulation with ALMA observations of Betelgeuse. For this figure, we have chosen an illustrative example of a simulation snapshot in time and orientation, where the nonrotating simulation can be easily mistaken as rotation. The left panels show the unconvolved simulation image ((a) and (d)). The middle panels show the convolved synthetic image ((b) and (e)). The right panels show the actual ALMA observations ((c) and (f)) from K18. The ALMA beam size is indicated by the gray circles. The top row demonstrates the continuum intensity maps. The measured radii $R_{\rm radio}$ of the radio photosphere are indicated by white circles. The bottom row shows the radial velocity maps measured from the Doppler shift of SiO lines. The green arrows indicate the rotational axes from the fit. Images from the simulation are rotated such that the rotational axis aligns with the observed axis. Credit for the right two panels: Kervella et al., A&A, 609, A67, 2018, reproduced with permission ESO. This figure is available as an animation over a real-time duration of 16 s and is also on Zenodo: doi:10.5281/zenodo.10199936. The animation shows the month-timescale variations in the intensity map and radial velocity map during a 5 yr long mock observation, where the time is marked in white in panel (a). The hot spot in the intensity map moves on the stellar disk or sometimes splits (panel (b)), and the radial velocity map sometimes appears multipolar (panel (e)).

(An animation of this figure is available.)

feature has also been observed in other images of Betelgeuse in the UV (Gilliland & Dupree 1996), optical (Young et al. 2000; Haubois et al. 2009; Montargès et al. 2016), and radio (O'Gorman et al. 2017). Quantitatively, the intensity in the mock image is half of the observed value. This corresponds to a discrepancy of 15% in temperature since the intensity is proportional to temperature to the fourth power. This is mostly due to the fact that the surface temperature is lower and the atmosphere is less extended than observed. For the radial velocity map, positive and negative radial velocity shifts partially cancel each other after convolution with the beam size. This blurs the original radial velocity map with convective motions up to 30 km s$^{-1}$ (Figure 2(d)) into a map with radial velocity variations up to 8 km s$^{-1}$ (Figure 2(e)) consistent with the ALMA observations. We conclude that the synthetic images of the intensity map and radial velocity qualitatively match the main features of the observations. In particular, both synthetic and observed radial velocity maps (Figures 2(e) and (f)) show a dipolar velocity field, with one hemisphere approaching the observer and the other receding.

K18 interpreted the dipolar velocity field as a sign of rotation. Fitting for this, assuming rigid rotation, they inferred a $v \sin i$ of $5.47 \pm 0.10$ km s$^{-1}$ and a residual velocity dispersion of 1.44 km s$^{-1}$. Following the same procedure for our synthetic image, we obtain very similar values; see Table 1. The resulting

**Table 1**
Comparison of Fitting Parameters of a Nonrotating RSG Simulation with ALMA Observations of Betelgeuse

| Inferred Parameters | $v \sin i$ (km s$^{-1}$) | $\sigma_{\rm res}$ (km s$^{-1}$) | $\chi^2_{\rm red}$ |
|---|---|---|---|
| Mock observation (this work, nonrotating) | $5.51 \pm 0.21$ | 1.68 | 94.6 |
| ALMA observation (K18) | $5.47 \pm 0.10$ | 1.44 | 20.7 |

**Note.** The radial velocity maps being fitted are shown in Figure 2. A rigidly rotating model is used to fit the radial velocity maps to obtain the perceived projected rotation rate $v \sin i$ and fitting parameters (standard deviation of the residual $\sigma_{\rm res}$ and reduced $\chi^2_{\rm red}$). All the parameters are obtained using the method described in Section 2.3.

$\chi^2_{\rm red}$ is larger than observed but comparable within an order of magnitude. As shown in Figure 5 later in the text, if we have underestimated the overall smearing effect of the observational pipeline, the $\chi^2_{\rm red}$ may be vastly decreased. In principle, this can be done in the future with the CASA package[8] to more properly take into account the antenna configuration, noises, etc.

---
[8] https://casa.nrao.edu

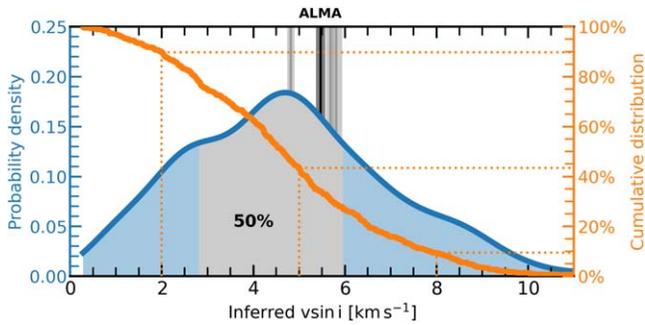

**Figure 3.** Probability distribution of apparent $v \sin i$ resulting from turbulent motions above the surface of a nonrotating stellar model. The gray vertical lines indicate the observed values based on SiO and CO lines (K18), with their uncertainties indicated by the widths and the SiO line we have modeled shown darker. The blue curve illustrates the kernel-smoothed probability density function of the inferred $v \sin i$. The 50% probability interval around the peak is shaded in gray. The orange curve shows the cumulative distribution function.

Note, however, that our simulation is based on a nonrotating star. The dipolar velocity field shown in the synthetic image is not the result of rotation but of the alignment of the group motions of the surface convective cells. The example shown here illustrates how turbulent motions in the atmosphere of a nonrotating RSG can, at certain times, mimic the effect of rotation. This raises the question of how confident we are that Betelgeuse is rotating and to what extent the expected large-scale convection affects the rotation measurement.

### 3.2. Probability of Inferring Rotation from the Radial Velocity Map of a Nonrotating RSG

In the previous section, we showed an example of how a nonrotating simulation can appear to be rotating as a result of underresolved convective motions. We had chosen a particular snapshot in time that illustrates this point well. In this section, we discuss how often such situations occur. We present the distribution of inferred $v \sin i$ in our mock observations for the simulated RSG. To obtain the probability distribution of inferred $v \sin i$ in a single-epoch observation, we compile 480 inferred $v \sin i$ calculated for six faces and 80 snapshots (uniformly taken across the 5 yr time span of the relaxed simulation) and plot the distribution in Figure 3.

Figure 3 shows that the apparent $v \sin i$ distribution inferred for our nonrotating models peaks at ∼5 km s$^{-1}$. The observed values for Betelgeuse (gray vertical lines) fall into the 50% probability interval around the peak. These simulations thus show that it is rather common that turbulent motions give rise to apparent $v \sin i$ similar to what is observed for Betelgeuse. We further show that for a nonrotating RSG to be interpreted as rotating faster than ∼5 km s$^{-1}$, the probability still remains as high as ∼40%. These numbers depend on the simulation adopted and the parameters assumed for Betelgeuse, which are not well constrained (Dolan et al. 2016; Joyce et al. 2020). Nevertheless, from the distribution, we expect that ∼90% single-epoch observations would show signs of rotation faster than ∼2 km s$^{-1}$. By visual inspection, the majority of those with inferred $v \sin i \lesssim 2$ km s$^{-1}$ do not show an obvious dipolar structure that could be mistaken for rotation. When analyzing an alternative simulation with slightly different stellar parameters (see Appendix D, in particular Figure 10), we also find a very high chance of inferring rotation at a few km s$^{-1}$.

### 4. Discussion

In Section 3, we argued that convective motions may be responsible for the dipolar velocity field observed for Betelgeuse as an alternative to the explanation of rapid rotation. The quantitative results we presented are subject to uncertainties due to the choices for the model parameters and limitations of simulations. We discuss this further in Appendices C.4 and D. Despite the uncertainties, we can use these results to formulate conceptual physical pictures that can be tested with future observations. In Figure 4, we illustrate three scenarios that, in principle, can explain the dipolar velocity map observed for the molecular shell around Betelgeuse.

1. Rapid rotation. This hypothesis, proposed by K18, states that Betelgeuse is rapidly rotating and drags the surrounding molecular shell along.
2. Large convective cells and stochastic effects. In this hypothesis, only very few convective cells cover the surface leading to stochastic effects (e.g., Schwarzschild 1975). At certain times, only two large cells may dominate the dynamics of the hemisphere. If one cell moves outward and one inward, and the motions are transported to the molecular shell via waves or shocks, this would result in a dipolar velocity field.
3. Smaller convective cells moving coherently. A dipolar velocity field may also arise when the convective cells are smaller but move semicoherently in groups. This physical picture most closely describes what we see in our simulations, where coherent motion is the result of deeper convective motions that operate over a length scale comparable to the stellar radius. Other mechanisms, e.g., nonradial oscillations (Lobel & Dupree 2001) as mentioned in K18, may also be able to drive such coherent motions.

Note that, in principle, the rotational velocity field and the turbulent motions can coexist in the molecular layer. The three hypotheses we list here emphasize which of the components dominate the velocity field.

### 4.1. Current Observational Constraints

All three hypotheses we consider can, in principle, reproduce the dipolar radial velocity map. Here we mention observational constraints that argue in favor of or against the scenarios depicted in Figure 4.

Support for the rotation hypothesis (1) has been claimed based on HST data probing the chromosphere of Betelgeuse (Uitenbroek et al. 1998; Harper & Brown 2006). They found an upward trend in the radial velocity while scanning over the surface from northwest to southeast and interpreted this as being the result of rotation. The magnitude and direction of the inferred rotation are consistent with the data taken by K18 almost 20 yr later.

This interpretation of the HST data is not without controversy. The radial velocities vary between observations taken a few months apart (see Figure 6 in Harper & Brown 2006). Other HST data analyzed by Lobel & Dupree (2001) instead show evidence for a reversal of velocities, and Jadlovský et al. (2023) found no sign of rotation. These findings are more consistent with hypotheses (2) and (3),

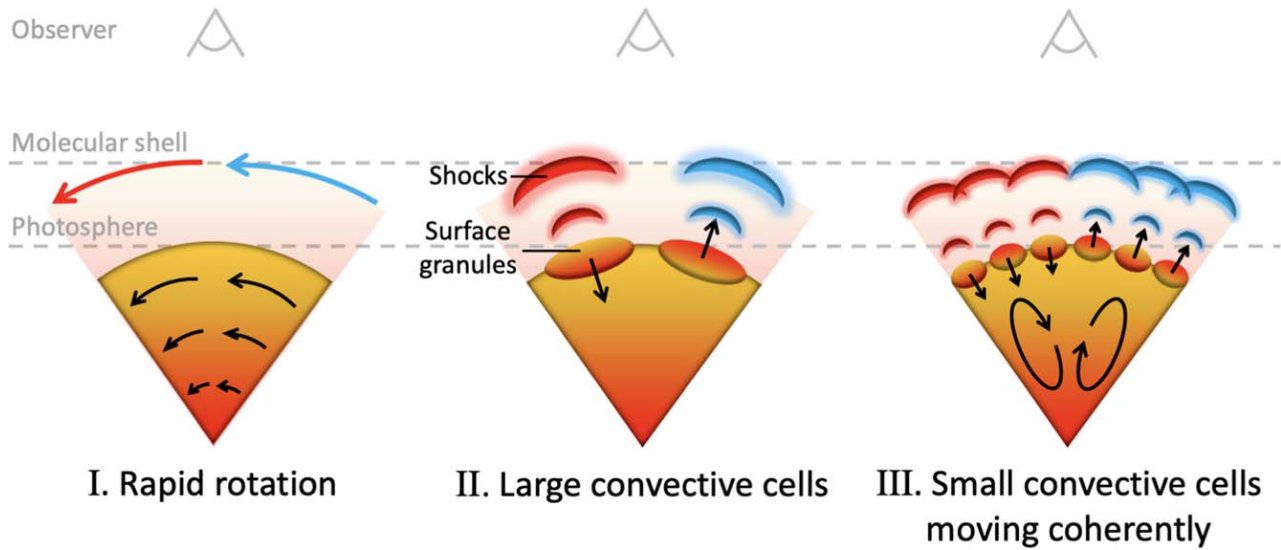

**Figure 4.** Three interpretations of Betelgeuse's dipolar radial velocity map observed by ALMA. The physical pictures and their observational consequences are explained in Section 4.

although it is unclear what the effect is of the great dimming event in the last study.

Evidence for the presence of (large) convective cells can be found in the asymmetries in the surface intensity map observed by ALMA but also in earlier data probing the UV, optical, and radio, as discussed in Section 3. Specifically, the infrared images taken with VLTI/MATISSE, which probe the bottom part of the molecular layer with a spatial resolution of 4 mas (Drevon et al. 2024), show multiple resolved structures. Other nearby RSGs such as Antares, V602 Car, and AZ Cyg show similar evidence (Ohnaka et al. 2017; Climent et al. 2020; Norris et al. 2021). However, so far Betelgeuse is the only RSG with a marginally resolved velocity map available from ALMA.

A further indication for turbulent motions at the molecular shell is the integrated line broadening observed by ALMA. The observed broadening of FWHM $24\,\mathrm{km\,s^{-1}}$ (Section 3.3.2 in K18) is significantly larger than what can be explained by the claimed rotation rate of $5.5\,\mathrm{km\,s^{-1}}$. This line width is consistent with what we predict in our simulations, about $18\,\mathrm{km\,s^{-1}}$ on average. There is thus evidence for additional broadening, likely originating from convective motions, consistent with hypotheses (2) and (3).

### 4.2. Future Observations to Test the Hypotheses

The hypotheses we propose have clear predictions that can be tested with future observations. In particular, additional epochs and higher spatial resolution are desired to resolve the variable radial velocity map.

Additional epochs will allow us to probe changes in the velocity field over time. In the case of rapid rotation (hypothesis 1), we expect that the radial velocity map will not change significantly. Both the magnitude of the rotation and alignment of the spin axis should be close to values obtained in 2015 by K18.

Instead, if the velocity field is dominated by turbulent motions, we expect the surface to readjust. Convective cells at the surface (hypothesis 2) are expected to change on timescales of a few months (see, e.g., Montargès et al. 2018; Norris et al. 2021, for RSG CE Tauri and AZ Cyg). Convective motions in the deep layers (hypothesis 3) may take years; see Appendix B for an order-of-magnitude estimate. We expect the surface velocity field to significantly change on these timescales. The field may still be dipolar but likely with a different orientation or may not display a dipolar feature. The error bars quoted by K18 for the projected rotational velocity ($0.1\,\mathrm{km\,s^{-1}}$, corresponding to a relative error of 2%) and the orientation angle of the rotation axis (3.5°, i.e., 1% of a full circle) are so small that the probability of inferring values within these error bars by chance in the future observations is negligible.

Observations of the radial velocity map at increased spatial resolution will also help distinguish the hypotheses we have outlined here. An increase by a factor of 2 should be feasible. For ALMA, higher resolution can, in principle, already be achieved by going to higher frequencies. Asaki et al. (2023) reported to have achieved a spatial resolution of 5 mas for a carbon-rich AGB star, which is 3 times better. When ALMA upgrades its current 16 km baseline to 32 km, we can expect an (additional) improvement of a factor of 2 (Carpenter et al. 2020).

In Figure 5, we present mock observations with different spatial resolutions. We expect that an increase of a factor of 2 in spatial resolution (panels (c) and (g)) will already marginally resolve the convective structure in both the radial velocity map and the intensity map and be able to distinguish between all three scenarios. On the contrary, decreasing the resolution by a factor of 2 ((a) and (e)) smears out any asymmetrical feature in the intensity map, and the radial velocity map is completely dominated by a dipolar pattern. As the spatial resolution of the interferometer increases (left to right), the measured radio photospheric radius monotonically decreases from $\sim$30 to $\sim$25 mas, while the peak magnitude of radial velocity monotonically increases from $\sim$6 to $\sim$30 $\mathrm{km\,s^{-1}}$. As a result of increasing radial velocity magnitude, the fitted $v\sin i$, $\chi^2_{\mathrm{red}}$, and $\sigma_{\mathrm{res}}$ all increase with higher resolution, among which $\chi^2_{\mathrm{red}}$ increases by 3 orders of magnitude. If future observations can be done with higher resolution, these are the clear signatures to look for to support our hypothesis, although the differences in the frequency range need to be considered.

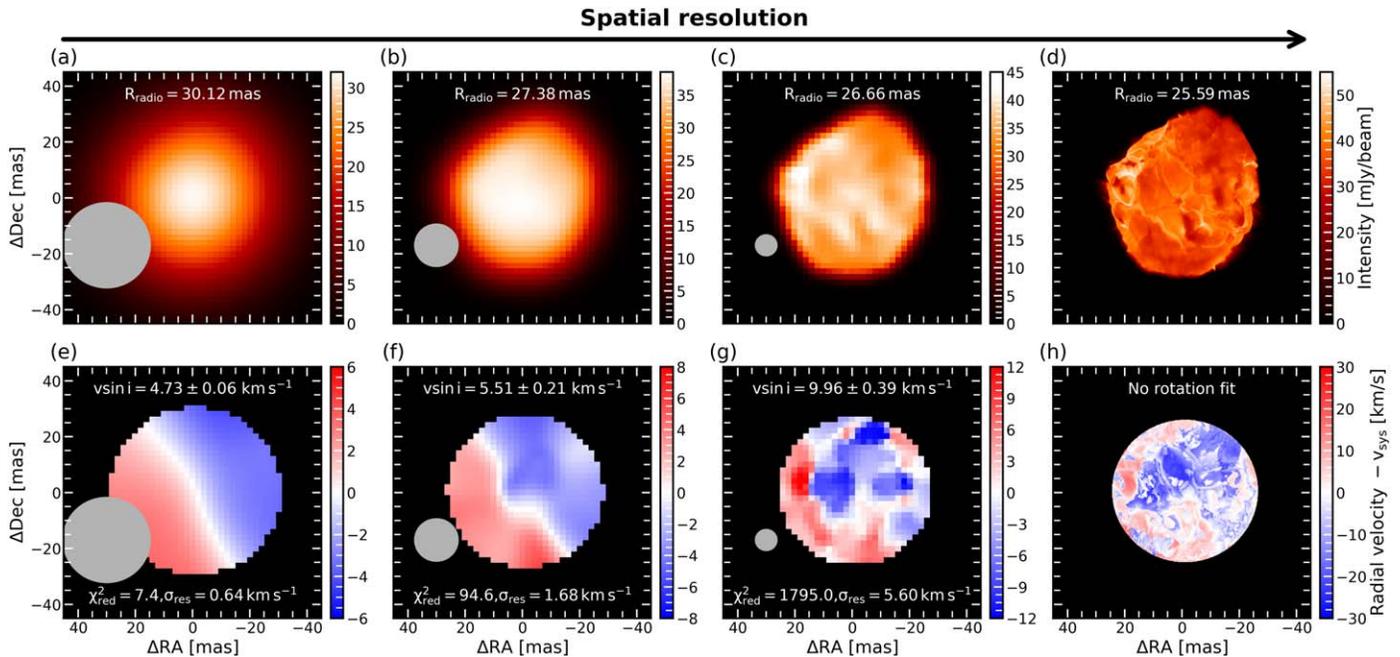

**Figure 5.** Effects of spatial resolution on the observed intensity map and radial velocity map in mock observations. From the left to the right, spatial resolution increases with decreasing beam size (indicated by gray circles). The second column ((b) and (f)) uses the standard resolution of FWHM = 18 mas adopted in ALMA observations of Betelgeuse (K18), the same as shown in Figure 2. In comparison, the beam size of the first column ((a) and (e)) is increased by a factor of 2 (36 mas), whereas the beam size of the third column ((c) and (g)) is reduced by a factor of 2 (9 mas). The last column ((d) and (h)) is the same original simulation image shown in Figure 2. Physical quantities marked in white are defined the same way as in Table 1 using the method in Section 2.3.

## 5. Conclusions

In this work, we investigate whether large-scale convection can affect the rotation measurement of Betelgeuse. We generate synthetic ALMA images from 3D CO5BOLD simulations of nonrotating RSGs and compare them to the actual ALMA observations of Betelgeuse (K18). Our conclusions are summarized as follows.

1. Large-scale motions in the RSG atmospheres generated by convection can be mistaken for rotation in interferometric observations. Both the convective cell size of the RSGs and the beam size used in ALMA observations are a large fraction of the stellar radius. Therefore, large-scale convection can be blurred as a dipolar feature in the radial velocity map that resembles rotation. This may apply to other cool stars, e.g., the ∼1 $M_\odot$ AGB star R Dor, which is reported to be rotating at $v \sin i \sim$ 1 km s$^{-1}$ with ALMA (Figure 1 in Vlemmings et al. 2018) but only displays turbulent motions in higher-resolution VLTI/AMBER observations probing similar heights above the photosphere (Figure 7 in Ohnaka et al. 2019). For unresolved lines in other types of stars, to what extent the turbulence complicates the spectroscopic $v \sin i$ determination is still not clear in both low-mass red giants (Patton et al. 2024) and massive stars (Simón-Díaz & Herrero 2014; Simón-Díaz et al. 2017; Schultz et al. 2023).
2. Future interferometric observations, e.g., with ALMA, are needed to assess the rotation of Betelgeuse. Our simulations suggest that another single-epoch observation of Betelgeuse with ALMA is sufficient to confirm if the observed maps show signs of rotation. Multi-epoch observations and higher spatial resolutions are desired for further constraints.[9]
3. The postprocessing package developed in this work can be applied to other forward modeling from 3D simulations of cool stars to synthetic radio spectra (e.g., Ramstedt et al. 2017; Doan et al. 2020; De Ceuster et al. 2022). This is particularly timely with the ongoing ALMA programs such as DEATHSTAR (Ramstedt et al. 2020; Andriantsaralaza et al. 2021) and ATOMIUM (Decin et al. 2020; Gottlieb et al. 2022; Decin et al. 2023; Montargès et al. 2023) that advance our understanding of chemistry, dust formation, planetary nebulae, binary interaction, and mass loss of these cool stars.
4. Regardless of whether Betelgeuse is rapidly rotating, more efforts are needed in both theoretical and observational aspects of RSGs. If Betelgeuse is indeed rapidly rotating, understanding the consequences of convection may well still be important for accurately interpreting the observational signatures of rotation. The stellar merger scenario for the rapid rotation (Wheeler et al. 2017; Chatzopoulos et al. 2020; Sullivan et al. 2020; Shiber et al. 2023) would demand further studies in the context of Betelgeuse's other properties, e.g., its runaway nature (Harper et al. 2008; Decin et al. 2012). If, on the other hand, future observations show evidence of turbulent motions, it may be possible for ALMA and other interferometers to trace the velocity field in different heights across the RSG atmosphere. Such observations would enable us to investigate the possible connections between pulsation, convection, and wind-launching mechanisms (Yoon & Cantiello 2010; Kee et al. 2021).

---

[9] Shortly after submission of this paper and posting it on the arXiv, a preliminary analysis of a new high-resolution ALMA image was presented during the conference "ALMA at 10 yr: Past, Present, and Future" (Dent 2023). Their data were taken in 2022 at higher frequency leading to a smaller beam size of approximately 8 mas (similar to our Figure 5 panels (c) and (g)). The radial velocity map was not yet available, but the team indicated that it is difficult to recognize a sign of rotation in their data. This is in line with our predictions, but we will need to await a full analysis of the data before conclusions can be drawn.

## Acknowledgments

We thank the referee for the dedicated comments that helped improve this work. We thank Pierre Kervella for sharing the processed data cubes and helpful discussions. We also acknowledge helpful discussions with Norbert Langer, Falk Herwig, and Leen Decin and many detailed comments from the community. A.C. acknowledges support from the French National Research Agency (ANR) funded project PEPPER (ANR-20-CE31-0002). B.F. has received funding from the European Research Council (ERC) under the European Union's Horizon 2020 research and innovation program (grant agreement No. 883867, project EXWINGS). The CO5BOLD simulations were enabled by resources provided by the Swedish National Infrastructure for Computing (SNIC) and PSMN in Lyon, France. This research project was partly conducted using computational resources (and/or scientific computing services) at the Max-Planck Computing & Data Facility. This research was supported by the Munich Institute for Astro-, Particle and BioPhysics (MIAPbP), which is funded by the Deutsche Forschungsgemeinschaft (DFG, German Research Foundation) under Germany's Excellence Strategy —EXC-2094—390783311.

*Software:* CO5BOLD (Freytag et al. 2012), FASTCHEM2 (Stock et al. 2018, 2022), MARCS (Gustafsson et al. 2008), MAGRITTE (De Ceuster et al. 2020a, 2020b, 2022), MESA (Paxton et al. 2011, 2013, 2015, 2018, 2019; Jermyn et al. 2023), Astropy (Astropy Collaboration et al. 2013, 2018, 2022), NumPy (Harris et al. 2020), SciPy (Virtanen et al. 2020), Matplotlib (Hunter 2007), Jupyter (Kluyver et al. 2016).

## Appendix A
## Single-star 1D Models

The fast rotation rate inferred for Betelgeuse by K18 is surprising in light of the predictions of single-star models (Wheeler et al. 2017; Chatzopoulos et al. 2020). We illustrated this in Figure 1 mentioned briefly in the main text. Here we provide additional details.

We show predictions for massive star evolutionary models. During their main-sequence evolution (core H burning, blue shading), massive stars rotate up to a few hundred km s$^{-1}$, but once they have expanded to become RSGs (core He burning, red shading), their rotation rates drop to 0.1–0.001 km s$^{-1}$. In general, the RSGs with higher initial masses rotate slower due to higher angular momentum loss via stronger stellar wind.

The inferred rotation rate for Betelgeuse is 2–3 orders of magnitude larger than what is predicted, violating the expectations for single-star evolution. The figure shows the measurement for Betelgeuse by K18 as a red box. The bottom of the box corresponds to an assumed orientation where we see the star equator on ($i = 90°$). The middle black line assumes $\langle \sin i \rangle = \pi/4$, which corresponds to the average for random orientations. The top of the box assumes $i = 20°$, which is the value originally suggested by Uitenbroek et al. (1998) based on the location of the hot spot.

The models shown are computed with 1D stellar evolution code MESA version r23.05.1 (Paxton et al. 2011, 2013, 2015, 2018, 2019; Jermyn et al. 2023), with initial masses of 12, 15, 18, and 21 $M_\odot$ and initial rotational velocities of 100, 200, and 300 km s$^{-1}$ at metallicity $Z = 0.02$. Convection is modeled according to mixing-length theory with $\alpha_{\rm MLT} = 1.5$ plus an exponential overshoot scheme (Herwig 2000) with coefficients $f_{\rm ov} = 0.0415$ and $f_{0,\rm ov} = 0.008$ (Brott et al. 2011). The convective boundary is determined with the Ledoux (1947) criterion, and we assume efficient semiconvection (Langer et al. 1983, 1985). During the main sequence, wind mass loss is accounted for as in Vink et al. (2001), while we switch to de Jager et al. (1988) when the effective temperature $T_{\rm eff}$ drops below $10^4$ K. We include the following mechanisms of rotational mixing and angular momentum transport (see Heger et al. 2000; Paxton et al. 2013): dynamical shear instability, secular shear instability, Eddington–Sweet circulation, Goldreich–Schubert–Fricke instability, and Spruit–Tayler dynamo (Spruit 2002; Heger et al. 2005; Paxton et al. 2013). The MESA inlists and history files are available at Zenodo: doi:10.5281/zenodo.10199936.

The tracks we show are subject to model uncertainties. For example, Beasor et al. (2020) argue that the wind prescription we used overestimates the mass-loss rates for RSGs, which would reduce angular momentum loss and spin down in the late phases. Another uncertainty is the mixing-length parameter. Goldberg et al. (2022) argue, based on 3D simulations, in favor of higher mixing-length parameters, which would lead to more compact giants. However, these uncertainties do not affect our main conclusion, because the the main reason for the significant spin-down is the large expansion that is needed for a main-sequence star to evolve and become a giant. We confirm this by running alternative MESA simulations where, for cool stars, we reduce the winds by a factor of 0.2 and gradually increase the mixing length parameter to 4. Even in such a conservative setup, the rotational velocities stay almost constant on the RSG branch between 0.01 and 0.3 km s$^{-1}$, still 2 orders of magnitude lower than the inferred rotation rate of Betelgeuse.

## Appendix B
## Order-of-magnitude Estimates

In this section, we estimate the properties of convection of RSGs with order-of-magnitude analytical arguments. The surface convective cell size is determined by the effective temperature $T_{\rm eff}$ and surface gravity $g$. The convective velocity, on the other hand, is determined by $T_{\rm eff}$ alone. These parameters can be estimated on the order-of-magnitude level, given our understanding of RSGs and measured properties of Betelgeuse. Note that in this section, the surface properties refer to the values at the infrared photosphere where the Rossland optical depth is of order unity, and not the molecular shell where ALMA probes. However, turbulent motions at the infrared photosphere are expected to propagate to the molecular shell through waves and shocks, thereby reflected in the ALMA observations.

Near the infrared photosphere, the horizontal size $d$ of the convective cell can be estimated from the pressure scale height $H_p$ multiplied by a factor $\alpha$ of order 10 for RSGs (Tremblay et al. 2013; Paladini et al. 2018). Therefore, the surface

convective cell size is

$$d \sim \alpha \frac{k_B T_{\rm eff}}{\mu m_u g} \sim 66\, R_\odot \left(\frac{\alpha}{10}\right)\left(\frac{T_{\rm eff}}{3600\,{\rm K}}\right)\left(\frac{g}{{\rm cm\,s^{-2}}}\right)^{-1}, \quad ({\rm B1})$$

where $k_B$ is the Boltzmann constant, $m_u$ is the proton mass, and $\mu$ is the mean molecular weight, taken to be 0.643 for near-solar composition. A similar method was also adopted to estimate the size of granules in 3D CO5BOLD RSG simulations (Chiavassa et al. 2011b). This estimate indicates that the surface convective cell size could be ~10% of the stellar radius within the uncertainties of parameters measured for Betelgeuse (Dolan et al. 2016; Joyce et al. 2020).

The convective velocity $v_c$ in the RSG envelope can be estimated by equating the energy flux to the convective flux,

$$v_c \sim \sqrt[3]{\frac{\sigma_{\rm SB} T_{\rm eff}^4}{\rho_s}} \sim 21\,{\rm km\,s^{-1}} \left(\frac{T_{\rm eff}}{3600\,{\rm K}}\right)^{\frac{4}{3}} \rho_{s,-9}^{-\frac{1}{3}}, \quad ({\rm B2})$$

where $\sigma_{\rm SB}$ is the Stefan–Boltzmann constant and $\rho_{s,-9}$ is the surface density in units of $10^{-9}\,{\rm g\,cm^{-3}}$ (representative density taken from the MESA models; see also Figure 2 of Goldberg et al. 2022). The constants of order unity are omitted here.

The timescale for the surface convective structure to readjust can be estimated as

$$t_{\rm surf} \sim \frac{d}{v_c} \sim 25\,{\rm days} \left(\frac{\alpha}{10}\right)\left(\frac{T_{\rm eff}}{3600\,{\rm K}}\right)^{-\frac{1}{3}}\left(\frac{g}{{\rm cm\,s^{-2}}}\right)^{-1} \rho_{s,-9}^{\frac{1}{3}}, \quad ({\rm B3})$$

typically on the order of weeks to months.

In comparison, the velocity field is predicted to be influenced by the deep convection, which operates over a larger length scale comparable to the stellar radius as shown in 3D simulations (Kravchenko et al. 2019). The timescale for the deep convection can thus be estimated as the convective turnover timescale,

$$t_{\rm turnover} \sim \frac{R}{v_c} \sim 0.8\,{\rm yr} \left(\frac{R}{800\,R_\odot}\right)\left(\frac{T_{\rm eff}}{3600\,{\rm K}}\right)^{-\frac{4}{3}} \rho_{s,-9}^{\frac{1}{3}}. \quad ({\rm B4})$$

## Appendix C
## Postprocessing Package: Methods, Tests, and limitations

### C.1. Equilibrium Chemistry with FASTCHEM2

The number densities of different species in our simulations are obtained from FASTCHEM2 assuming chemical equilibrium. On-the-fly calculations are time-consuming given the large number of grid points and snapshots in the simulations. Therefore, we precalculate the number density tables that cover the parameter space of $2.5 \leqslant \log_{10}(T/[{\rm K}]) \leqslant 5.5$, $-30 \leqslant \log_{10}(\mathcal{R}) \leqslant -9$, where $\mathcal{R} = (P/[{\rm dyn\,cm^{-2}}])/(T/[{\rm K}])^4$, using FASTCHEM2. Here, $T$ is the gas temperature and $P$ is the gas pressure. The number densities in our simulations are interpolated values from these tables. To verify that the interpolated results from the precalculated FASTCHEM2 table are suitable for RSG simulations, we compare the interpolated values to the values given by a MARCS RSG model in Figure 6. The MARCS code was developed to create 1D atmospheric models for evolved stars (Gustafsson et al. 2008), which have been widely used to produce spectra to obtain basic parameters of RSGs from spectroscopy (e.g., Levesque et al. 2005). The interpolated FASTCHEM2 values closely agree with the MARCS values.

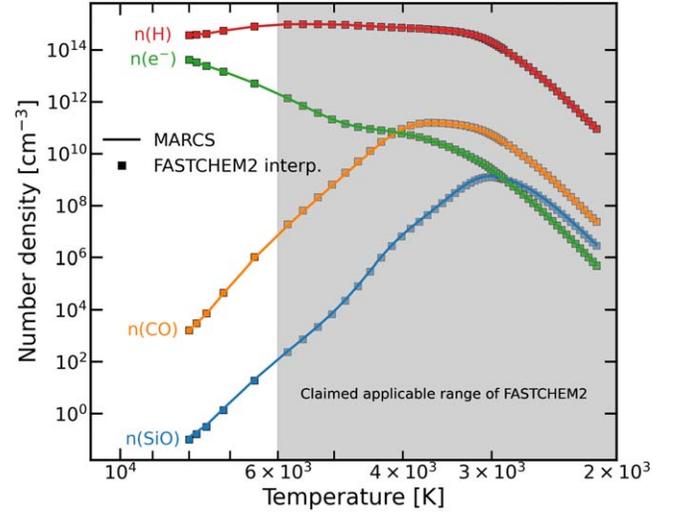

**Figure 6.** Comparison of number densities from interpolated results of the precalculated FASTCHEM2 table (squares) against those from MARCS (curves). The colors indicate the number densities of atomic hydrogen, electrons, CO molecules, and SiO molecules.

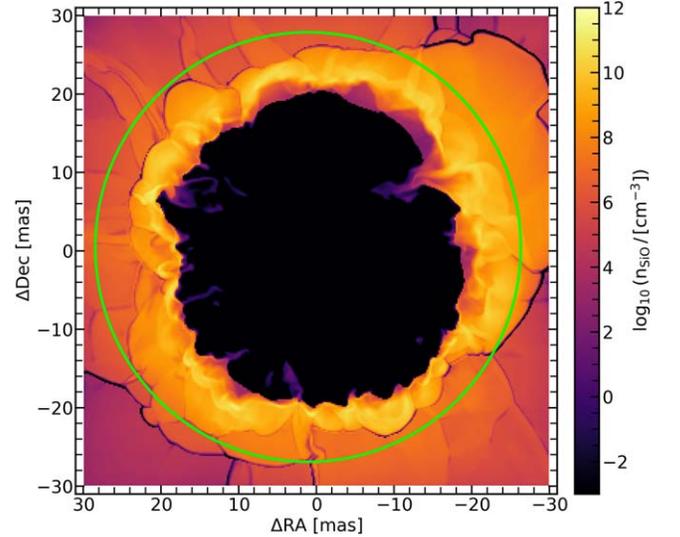

**Figure 7.** 2D middle slice of SiO number density in the simulated RSG envelope. The green circle is the radio photosphere of the star. From the same snapshot of the simulation shown in Figure 2.

In Figure 7, we show an example of the SiO number density distribution in our simulations by taking a 2D slice along the middle cross section. The green circle indicates the radio photosphere determined by the continuum intensity map (see Section 2.2). In our simulations, molecules are dissociated in the inner envelope due to high temperature. The most abundant SiO molecules are found near the infrared photosphere due to recombination. The presence of shocks outside of the photosphere can be seen from the SiO number density slice.

### C.2. Continuum and Line Radiative Transfer

In our postprocessing, we numerically integrate the time-independent radiative transfer equation along the ray neglecting scattering. A distance $x$ can be defined from the origin of the ray. The optical depth is

$$\tau_\nu(\hat{\boldsymbol{n}}, \boldsymbol{r}) \equiv \int_0^{x(\hat{\boldsymbol{n}},\boldsymbol{r})} \chi_\nu dx'. \quad (C1)$$

Here, $\hat{\boldsymbol{n}}$ is the unit vector along the ray direction, and $\boldsymbol{r}$ is the vector of position. The intensity can then be calculated as

$$I_\nu(x) = I_\nu(0)e^{-\tau_\nu} + \int_{e^{-\tau_\nu}}^{1} S_\nu(\tau_\nu - \tau'_\nu)d(e^{-\tau'_\nu}), \quad (C2)$$

where $S_\nu$ is the source function defined by $S_\nu \equiv \eta_\nu/\chi_\nu$. $I_\nu$, $\eta_\nu$, and $\chi_\nu$ are the intensity, emissivity, and opacity at frequency $\nu$. The intensity $I_\nu(0)$ at the origin is taken to be 0.

For the wavelength range studied here, the emissivity and opacity are contributed by continuum emission and molecular lines, namely, $\eta_\nu = \eta_\nu^{\rm con} + \eta_\nu^{\rm line}$, $\chi_\nu = \chi_\nu^{\rm con} + \chi_\nu^{\rm line}$. We assume that $\eta_\nu^{\rm con} = \chi_\nu^{\rm con} B_\nu$, where $B_\nu$ is the Planck function. The continuum opacity $\chi_\nu^{\rm con}$, dominated by $H^-$ free–free transitions, is taken from the analytical fits in Harper et al. (2001) specifically made for radio observations of Betelgeuse. Relevant quantities for $^{28}$SiO ($\nu = 2$, $J = 8-7$) lines are obtained from the ExoMol database[10] (Tennyson & Yurchenko 2012; Tennyson et al. 2016) using the results of Yurchenko et al. (2022) and further computed assuming local thermal equilibrium (LTE). For this transition, we did not find available information for the collisional rates; therefore, we refrained from performing detailed non-LTE calculations. The line emissivity and opacity contributed by the transition from level $i$ to level $j$ are computed following standard assumptions:

$$\eta_\nu^{{\rm line},ij} = \frac{h\nu}{4\pi} n_i A_{ij} \phi_\nu^{ij}, \quad (C3)$$

$$\chi_\nu^{{\rm line},ij} = \frac{h\nu}{4\pi} (n_j B_{ji} - n_i B_{ij}) \phi_\nu^{ij}, \quad (C4)$$

where $h$ is the Planck constant, $n_i$ is the number density of the species at level $i$, and $A_{ij}$, $B_{ij}$ are Einstein coefficients. The line profile function is assumed to be a Gaussian with a line width contributed by thermal broadening,

$$\phi_\nu^{ij} = \frac{1}{\delta\nu_{ij}\sqrt{\pi}} \exp\left[-\left(\frac{\nu - \tilde{\nu}_{ij}}{\delta\nu_{ij}}\right)^2\right], \quad (C5)$$

$$\delta\nu_{ij} = \frac{|v_{\rm therm}|}{c} \tilde{\nu}_{ij}, \quad (C6)$$

where the Doppler-shifted frequency $\tilde{\nu}_{ij}$ and the thermal velocity $v_{\rm therm}$ are

$$\tilde{\nu}_{ij} = \nu_{ij}\left(1 - \frac{\hat{\boldsymbol{n}} \cdot \boldsymbol{v}}{c}\right),$$

$$v_{\rm therm} = \sqrt{\frac{2k_B T}{m}}. \quad (C7)$$

Here, $\nu_{ij}$ is the rest frequency for the transition, $\boldsymbol{v}$ is the local velocity, $c$ is the speed of light, $T$ is the gas temperature, and $m$ is the particle mass of the species. The Doppler shift follows the convention that velocity toward the observer is blueshifted and has negative values.

We have tested our algorithm against the line transfer code MAGRITTE (De Ceuster et al. 2020a, 2020b, 2022) by applying both of them to the CO5BOLD simulation snapshot. The results agree with each other. However, MAGRITTE needs to construct the Voronoi grid. In comparison, our method runs faster with CO5BOLD snapshots that are computed on a Cartesian grid and does not suffer from extra interpolation errors. Additionally, MAGRITTE has not yet incorporated continuum opacity or interface to the ExoMol database, both of which are important for this study.

---

[10] https://www.exomol.com

### C.3. Gaussian Line Fit and Rotation Measurement

An example of our Gaussian line fit procedure is illustrated in Figure 8. Green arrows connect the line profiles to their corresponding pixels (indicated in black). To obtain the radial velocity map, we fit a Gaussian (orange solid line) to the line profile (blue solid line) in each pixel of the image and find the mean value of the Gaussian (orange dotted line).

Following Section 3.5 in K18, we subtract the inferred systematic velocity from the radial velocity map and fit a projected radial velocity map of a rigidly rotating sphere to the synthesized map. Namely, we assume that the $v_{\rm sys}$-subtracted radial velocity map has a velocity distribution of $v(x, y) = v \sin i \times (x\cos\theta + y\sin\theta)/R_{\rm radio}$, where the fitted parameters are the position angle $\theta$ and the inferred $v \sin i$, and $x$, $y$ are the coordinates of the 2D radial velocity map.

We caution that there are different radii involved in interpreting the observations and comparing with 1D or 3D models: the infrared photosphere $R_{\rm infrared}$ approximately used as outer boundaries in 1D models, the radio photosphere $R_{\rm radio}$ obtained from ALMA continuum intensity maps, and the molecular shell radius $R_{\rm shell}$ probed through molecular lines in ALMA. It generally follows that $R_{\rm infrared} < R_{\rm radio} < R_{\rm shell}$, and for Betelgeuse, each two nearby radii are different by a factor of $\sim$1.3 (K18). Instead of using the radio photospheric radius $R_{\rm radio}$ for the fit, we suggest it is more consistent with the physical scenario to use the radius $R_{\rm shell} \approx 1.3 R_{\rm radio}$ of the molecular shell probed by ALMA. This would increase the measured $v \sin i$ at the molecular shell by a factor of 1.3, i.e., $v \sin i_{\rm shell} \approx 7$ km s$^{-1}$ at the molecular shell. Assuming that the molecular shell corotates with the radio photosphere, this translates to a rotational velocity at the radio photosphere of a factor of 1.3 less than the value measured at the molecular shell, i.e., $v \sin i_{\rm radio} \approx 5.5$ km s$^{-1}$ at the radio photosphere. However, if the molecular shell and the radio photosphere are not fully coupled, e.g., assuming a similar specific angular momentum between the molecular shell and the radio photosphere, then the rotational velocity at the radio photosphere would be another factor of 1.3 higher, namely, $v \sin i_{\rm radio} \approx 9$ km s$^{-1}$ at the radio photosphere. Since the infrared photosphere $R_{\rm infrared}$ is yet another factor of 1.3 smaller than the radio photosphere $R_{\rm radio}$, the measurement by K18 would give $v \sin i_{\rm infrared} \approx 4$ km s$^{-1}$ at the infrared photosphere $R_{\rm infrared}$ assuming constant angular frequency (corotation) outside the star, or $v \sin i_{\rm infrared} \approx 12$ km s$^{-1}$ assuming constant specific angular momentum outside the star. In this work, we closely follow the procedure done by K18 for direct comparison and thus use $R_{\rm radio}$ for the rotation fit, with the underlying assumption that the molecular shell corotates with the radio photosphere.

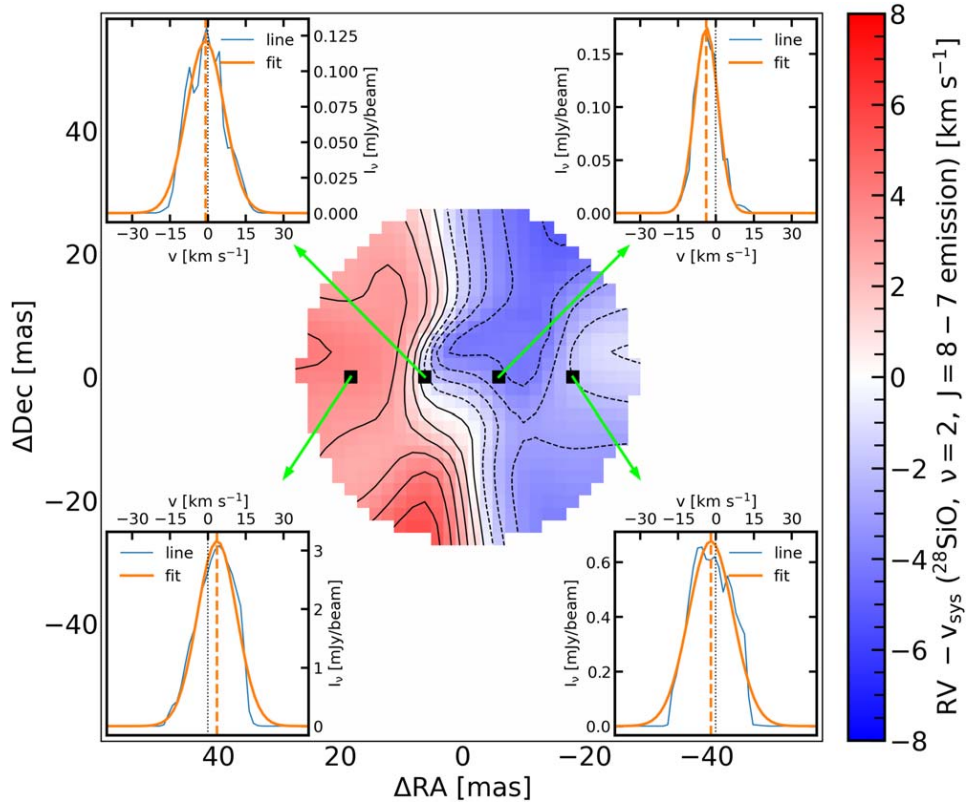

**Figure 8.** Examples of emission line profiles from synthetic ALMA images. Four small subplots show the simulated line profiles (blue) and fitted Gaussian (orange) in four pixels (shaded in black) of the synthetic ALMA image. In each pixel, the mean value (orange dashed vertical line) of the fitted Gaussian is taken to be the radial velocity from the Doppler shift. From the same snapshot of the simulation shown in Figure 2.

### C.4. Discrepancies from Observations and Limitations

We place the simulated star at a distance away from the observers such that it has an inferred radio photospheric radius $R_{radio}$ of ∼30 mas. However, this corresponds to a distance of 130 pc, consistent with the original Hipparcos parallax ($131^{+35}_{-23}$ pc; ESA 1997; Perryman et al. 1997) and its revised value ($152^{+22}_{-17}$ pc; Van Leeuwen 2007) but much closer than the distance derived from Hipparcos data combined with radio observations ($197^{+55}_{-35}$ pc; Harper et al. 2008) and its revised value ($222^{+48}_{-34}$ pc; Harper et al. 2017). As a comparison, asteroseismic constraints suggested a medium value ($168^{+27}_{-15}$ pc; Joyce et al. 2020). This discrepancy between the value adopted in the synthetic images and the observed value is partly because the fiducial simulation used in this work has a smaller radius than Betelgeuse and therefore needs to be placed closer to get a similar parallax. Another reason is rooted in the limitations of the 3D models, which have been shown to be less extended than actual RSGs (Arroyo-Torres et al. 2015; Climent et al. 2020; Chiavassa et al. 2022). This could be due to the fact that the radiation pressure is not included in the simulations (Chiavassa 2022).

The simulated continuum intensity is about half of the observed value (see Figure 2), and the line amplitude is an order of magnitude lower than observed (see Figure 8 compared to Figures 1 and 9 in K18). This could result from the small atmospheric extension of the simulations or not including nonequilibrium chemistry, non-LTE populations, dust emission, maser, or scattering in the postprocessing. However, as we are interested in the radial velocity shifts instead of the molecular abundances in this study, the effects of these missing physics are expected to be limited. This is supported by our simulations, which show no significant height dependence in the velocity field from the surface to the molecular layer. From the physical point of view discussed in Appendix B, the convective velocity $v_c$ scales very weakly to the surface energy flux $F$ and surface density $\rho_s$ as $(F/\rho_s)^{1/3}$, so the magnitude of the convective velocity is only weakly affected by the limitations of simulations.

A more uncertain aspect is whether the convective structure in our simulations can represent actual RSG convection. The convective length scales are subject to the uncertainties of the stellar parameters of Betelgeuse itself, as well as the resolution of the simulations (Chiavassa et al. 2011a). Aside from that, core convection simulations suggest that large-scale convection is dominated by dipolar flows (see, e.g., Lecoanet & Edelmann 2023, for a review). For typical RSGs, where the bottom of the convective envelope is less than 0.05 of the stellar radius, we expect that similar dipolar flows may also dominate in the convective envelope. These dipolar flows are partly damped in our simulations by the drag forces in the central region (Freytag et al. 2012). However, we expect that an enhancement in the dipolar flows would only make our argument stronger.

### Appendix D
### Additional Simulations

The parameters of two sets of 3D simulations used in this study are shown in Table 2. The fiducial simulation is

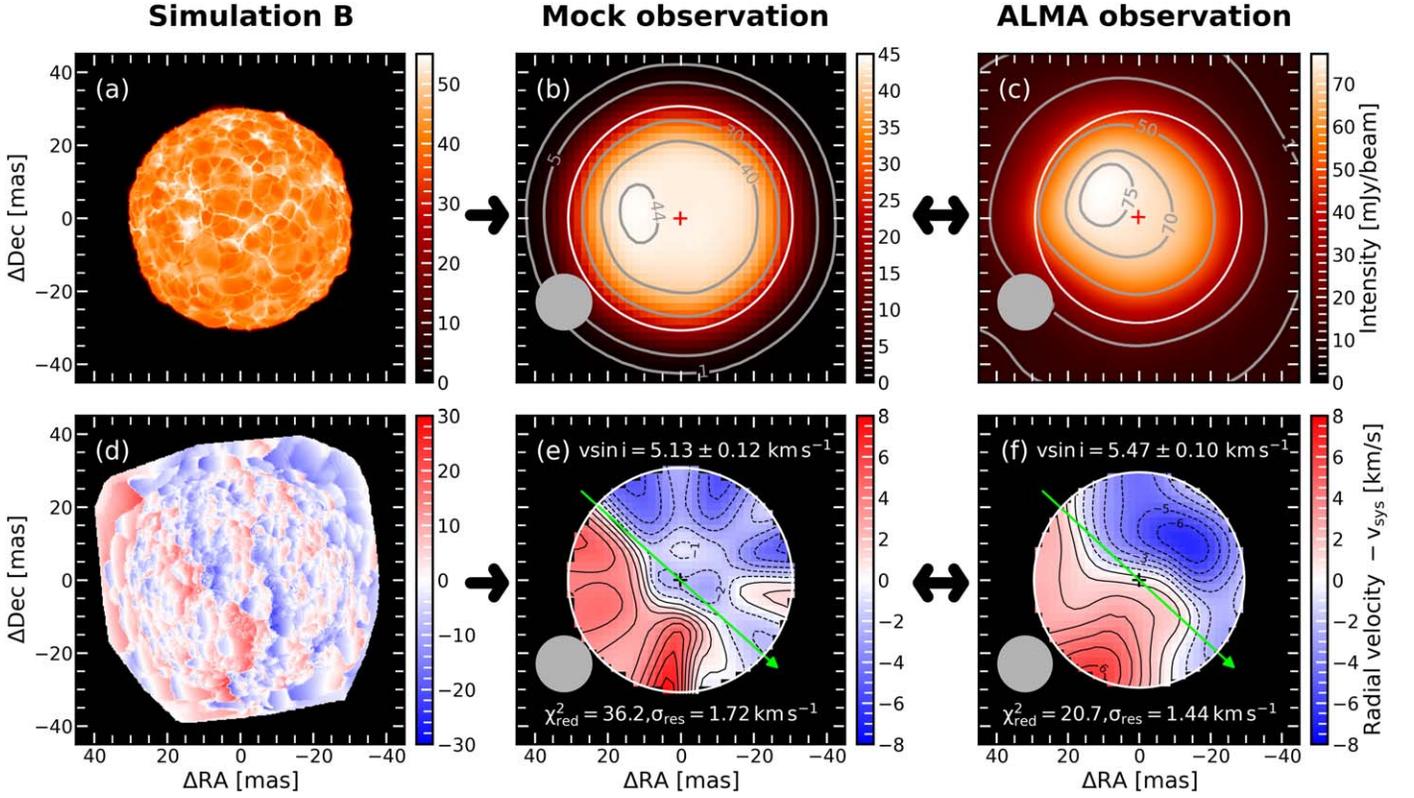

**Figure 9.** Continuum intensity map and radial velocity map of the best-fit simulation B snapshot, compared with ALMA observations of Betelgeuse. Physical quantities marked in white are defined the same way as in Table 1 using the method in Section 2.3. Credit for the right two panels: Kervella et al., A&A, 609, A67, 2018, reproduced with permission ESO.

**Table 2**
Parameters of the CO5BOLD Simulations Taken from Ahmad et al. (2023) Compared to the Observed Values for Betelgeuse

| | | $\log_{10} g$ (cm s$^{-2}$) | $T_{\rm eff}$ (K) | $\log_{10} L_\star$ ($L_\odot$) | $R_\star$ ($R_\odot$) | $M_{\rm pot}$ ($M_\odot$) | $M_{\rm env}$ ($M_\odot$) | $t_{\rm turnover}$ (yr) | Grid (No.) | $x_{\rm box}$ ($R_\odot$) | Snapshot (No.) | $t_{\rm snap}$ (yr) |
|---|---|---|---|---|---|---|---|---|---|---|---|---|
| Sim. | Fiducial (A) (st35gm04n37) | −0.419 | 3366 | 4.62 | 597 | 5.0 | 0.5 | ∼0.7 | $315^3$ | 1626 | 80 | 5.1 |
| | Alternative (B) (st35gm03n020) | −0.246 | 3620 | 4.95 | 759 | 12.0 | 3.0 | ∼0.8 | $637^3$ | 2093 | 100 | 6.3 |
| Obs. | Betelgeuse | $-0.5^{\rm a}$ or $0.0^{+0.3\,\rm b}_{-0.3}$ | $3500^{+200\,\rm c}_{-200}$ | $5.10^{+0.22\,\rm d}_{-0.22}$ | $764^{+116\,\rm e}_{-62}$ | ... | ... | ... | ... | ... | ... | ... |

**Notes.** The stellar parameters for both simulations and observations are listed as surface gravity $\log_{10} g$, effective temperature $T_{\rm eff}$, luminosity $\log_{10} L_\star$, and radius $R_\star$. All these surface quantities in simulations are taken at the local minimum point of specific entropy averaged over spherical shells and time, as described in Ahmad et al. (2023). We also list the mass $M_{\rm pot}$ used for the gravitational potential and envelope mass $M_{\rm env}$ for the simulations. The convective turnover timescale $t_{\rm turnover}$ is estimated analytically as $R_\star/v_c$, where the convective velocity $v_c$ is calculated using the equations in Appendix B. The number of grid cells, the simulation box size $x_{\rm box}$, the number of snapshots, and the physical time $t_{\rm snap}$ covered by the extracted snapshots are listed in sequence.
[a] Lobel & Dupree (2000).
[b] Lambert et al. (1984).
[c] Dolan et al. (2016; chosen value motivated by multiple sources).
[d] Harper et al. (2008).
[e] Joyce et al. (2020).

discussed in the main text (Figures 2 and 3), while an alternative simulation with smaller surface granules is presented in Figures 9 and 10. Due to the small granule size compared to the ALMA beam size, the radial velocity map displays small-scale features (Figure 9) and therefore is less likely to be mistaken for rotation (Figure 10). The change in the intensity map, however, is less obvious. In particular, its radial velocity map shows a velocity magnitude similar to the fiducial simulation but with more cell-like structures. This indicates that the magnitude of velocity in the radial velocity map may only be determined by the maximum convective velocity and the relative beam size and is not sensitive to the cell size. The morphology of the radial velocity map, however, is sensitive to the cell size and, consequently, to the surface gravity.

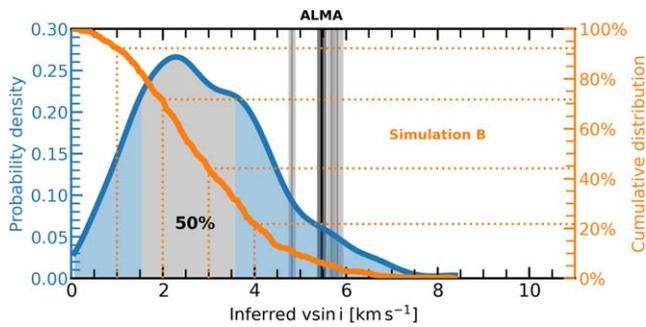

**Figure 10.** Probability distribution of apparent $v \sin i$, similar to Figure 3 but inferred from simulation B.

## ORCID iDs

Jing-Ze Ma (马竟泽) ● https://orcid.org/0000-0002-9911-8767
Andrea Chiavassa ● https://orcid.org/0000-0003-3891-7554
Selma E. de Mink ● https://orcid.org/0000-0001-9336-2825
Ruggero Valli ● https://orcid.org/0000-0003-3456-3349
Stephen Justham ● https://orcid.org/0000-0001-7969-1569
Bernd Freytag ● https://orcid.org/0000-0003-1225-8212